\PassOptionsToPackage{unicode}{hyperref}
\PassOptionsToPackage{hyphens}{url}
\PassOptionsToPackage{dvipsnames,svgnames,x11names}{xcolor}
\documentclass[
  article,
  nojss]{jss}

\usepackage{amsmath,amssymb}
\usepackage{iftex}
\ifPDFTeX
  \usepackage[T1]{fontenc}
  \usepackage[utf8]{inputenc}
  \usepackage{textcomp} 
\else 
  \usepackage{unicode-math}
  \defaultfontfeatures{Scale=MatchLowercase}
  \defaultfontfeatures[\rmfamily]{Ligatures=TeX,Scale=1}
\fi
\usepackage{lmodern}
\ifPDFTeX\else  
\fi
\IfFileExists{upquote.sty}{\usepackage{upquote}}{}
\IfFileExists{microtype.sty}{
  \usepackage[]{microtype}
  \UseMicrotypeSet[protrusion]{basicmath} 
}{}
\makeatletter
\@ifundefined{KOMAClassName}{
  \IfFileExists{parskip.sty}{%
    \usepackage{parskip}
  }{
    \setlength{\parindent}{0pt}
    \setlength{\parskip}{6pt plus 2pt minus 1pt}}
}{
  \KOMAoptions{parskip=half}}
\makeatother
\usepackage{xcolor}
\makeatletter
\ifx\paragraph\undefined\else
  \let\oldparagraph\paragraph
  \renewcommand{\paragraph}{
    \@ifstar
      \xxxParagraphStar
      \xxxParagraphNoStar
  }
  \newcommand{\xxxParagraphStar}[1]{\oldparagraph*{#1}\mbox{}}
  \newcommand{\xxxParagraphNoStar}[1]{\oldparagraph{#1}\mbox{}}
\fi
\ifx\subparagraph\undefined\else
  \let\oldsubparagraph\subparagraph
  \renewcommand{\subparagraph}{
    \@ifstar
      \xxxSubParagraphStar
      \xxxSubParagraphNoStar
  }
  \newcommand{\xxxSubParagraphStar}[1]{\oldsubparagraph*{#1}\mbox{}}
  \newcommand{\xxxSubParagraphNoStar}[1]{\oldsubparagraph{#1}\mbox{}}
\fi
\makeatother

\providecommand{\tightlist}{%
  \setlength{\itemsep}{0pt}\setlength{\parskip}{0pt}}\usepackage{longtable,booktabs,array}
\usepackage{calc} 
\usepackage{etoolbox}
\makeatletter
\patchcmd\longtable{\par}{\if@noskipsec\mbox{}\fi\par}{}{}
\makeatother
\IfFileExists{footnotehyper.sty}{\usepackage{footnotehyper}}{\usepackage{footnote}}
\makesavenoteenv{longtable}
\usepackage{graphicx}
\makeatletter
\def\maxwidth{\ifdim\Gin@nat@width>\linewidth\linewidth\else\Gin@nat@width\fi}
\def\maxheight{\ifdim\Gin@nat@height>\textheight\textheight\else\Gin@nat@height\fi}
\makeatother
\setkeys{Gin}{width=\maxwidth,height=\maxheight,keepaspectratio}
\makeatletter
\def\fps@figure{htbp}
\makeatother

\usepackage{orcidlink,thumbpdf,lmodern}

\newcommand{\class}[1]{`\code{#1}'}
\newcommand{\fct}[1]{\code{#1()}}
\usepackage{float}
\makeatletter
\@ifpackageloaded{caption}{}{\usepackage{caption}}
\AtBeginDocument{%
\ifdefined\contentsname
  \renewcommand*\contentsname{Table of contents}
\else
  \newcommand\contentsname{Table of contents}
\fi
\ifdefined\listfigurename
  \renewcommand*\listfigurename{List of Figures}
\else
  \newcommand\listfigurename{List of Figures}
\fi
\ifdefined\listtablename
  \renewcommand*\listtablename{List of Tables}
\else
  \newcommand\listtablename{List of Tables}
\fi
\ifdefined\figurename
  \renewcommand*\figurename{Figure}
\else
  \newcommand\figurename{Figure}
\fi
\ifdefined\tablename
  \renewcommand*\tablename{Table}
\else
  \newcommand\tablename{Table}
\fi
}
\@ifpackageloaded{float}{}{\usepackage{float}}
\floatstyle{ruled}
\@ifundefined{c@chapter}{\newfloat{codelisting}{h}{lop}}{\newfloat{codelisting}{h}{lop}[chapter]}
\floatname{codelisting}{Listing}

\makeatother
\makeatletter
\@ifpackageloaded{caption}{}{\usepackage{caption}}
\@ifpackageloaded{subcaption}{}{\usepackage{subcaption}}
\makeatother
\makeatletter
\@ifpackageloaded{tcolorbox}{}{\usepackage[skins,breakable]{tcolorbox}}
\makeatother
\makeatletter
\@ifundefined{shadecolor}{\definecolor{shadecolor}{rgb}{.97, .97, .97}}{}
\makeatother
\makeatletter
\ifdefined\Shaded\renewenvironment{Shaded}{\begin{tcolorbox}[sharp corners, boxrule=0pt, enhanced, interior hidden, borderline west={3pt}{0pt}{shadecolor}, breakable, frame hidden]}{\end{tcolorbox}}\fi
\makeatother

\ifLuaTeX
  \usepackage{selnolig}  
\fi
\usepackage{bookmark}

\IfFileExists{xurl.sty}{\usepackage{xurl}}{} 
\hypersetup{
  pdftitle={: Sparse Projected Averaged Regression in },
  pdfauthor={Roman Parzer; Laura Vana Gür; Peter Filzmoser},
  pdfkeywords={random projection, variable screening, ensemble
learning, R},
  colorlinks=true,
  linkcolor={blue},
  filecolor={Maroon},
  citecolor={Blue},
  urlcolor={Blue},
  pdfcreator={LaTeX via pandoc}}

\author{Roman Parzer~\orcidlink{0000-0003-0893-3190}\\TU Wien \And Laura
Vana Gür~\orcidlink{0000-0002-9613-7604}\\TU Wien \AND Peter
Filzmoser~\orcidlink{0000-0002-8014-4682}\\TU Wien}
\Plainauthor{Roman Parzer, Laura Vana Gür, Peter
Filzmoser} 

\title{\pkg{spar}: Sparse Projected Averaged Regression in \proglang{R}}
\Plaintitle{: Sparse Projected Averaged Regression
in} 

\Abstract{Package \pkg{spar} for \proglang{R} builds ensembles of
predictive generalized linear models with high-dimensional predictors.
It employs an algorithm utilizing variable screening and random
projection tools to efficiently handle the computational challenges
associated with large sets of predictors. The package is designed with a
strong focus on extensibility. Screening and random projection
techniques are implemented as \proglang{S}3 classes with user-friendly
constructor functions, enabling users to easily integrate and develop
new procedures. This design enhances the package's adaptability and
makes it a powerful tool for a variety of high-dimensional
applications.}

\Plainkeywords{random projection, variable screening, ensemble
learning, R}

\Address{
Roman Parzer\\
Computational Statistics (CSTAT), Institute of Statistics and
Mathematical Methods in Economics\\
Wiedner Hauptstraße 8-10\\
Vienna Austria\\
E-mail: \email{romanparzer1@gmail.com}\\
\\~
Laura Vana Gür\\
Computational Statistics (CSTAT), Institute of Statistics and
Mathematical Methods in Economics\\
Wiedner Hauptstraße 8-10\\
Vienna Austria\\
\\~
Peter Filzmoser\\
Computational Statistics (CSTAT), Institute of Statistics and
Mathematical Methods in Economics\\
Wiedner Hauptstraße 8-10\\
Vienna Austria\\
\\~

}

\begin{document}
\maketitle

\section{Introduction}\label{sec-intro}

The \pkg{spar} package for \proglang{R} \citep{RLanguage} offers
functionality for estimating generalized linear models (GLMs) in
high-dimensional settings, where the number of predictors \(p\)
significantly exceeds the number of observations \(n\) i.e., \(p>n\) or
even \(p\gg n\). To address the challenges of high dimensionality, the
package implements an algorithm which integrates variable screening
methods with random projection techniques to effectively reduce the
predictor space.

Random projection is a computationally-efficient method which linearly
maps a set of points in high dimensions into a much lower-dimensional
space. Several packages in \proglang{R} provide functionality for random
projections. For instance, package \pkg{RandPro}
\citep{RandProR, SIDDHARTH2020100629} allows a Gaussian random matrix, a
sparse matrix \citep{ACHLIOPTAS2003JL, LiHastie2006VerySparseRP} or a
matrix generated using the equal probability distribution with the
elements \(\{-1,1\}\) to be applied to the predictor matrix before
employing one of \(k\) nearest neighbor, support vector machine or naive
Bayes classifier on the projected features. Package \pkg{SPCAvRP}
\citep{SPCAvRPR} implements sparse principal component analysis, based
on the aggregation of eigenvector information from
``carefully-selected'' axis-aligned random projections of the sample
covariance matrix. Additionally, package \pkg{RPEnsembleR}
\citep{RPEnsembleR} implements a similar idea when building the ensemble
of classifiers: for each classifier in the ensemble, a collection of
(Gaussian, axis-aligned projections, or Haar) random projection matrices
is generated, and the one that minimizes a risk measure for
classification on a test set is selected. For \proglang{Python}
\citep{Python} the \pkg{sklearn.random\_projection} module
\citep{pedregosa2011scikit} implements two types of unstructured random
matrices, namely Gaussian random matrix and sparse random matrix.

Random projection can suffer from noise accumulation for very large
\(p\), as too many irrelevant predictors are being considered for
prediction purposes \citep{Dunson2020TargRandProj}. Therefore, screening
out irrelevant variables before performing the random projection is
advisable in order to tackle this issue. The screening can be performed
in a probabilistic fashion, by randomly sampling covariates for
inclusion in the model based on probabilities proportional to an
importance measure (as opposed to random subspace sampling employed in,
e.g., random forests). The \proglang{R} landscape for variable screening
techniques is very rich. An overview of some notable packages on the
Comprehensive \proglang{R} Archive Network (CRAN) includes the following
packages. Package \pkg{SIS} \citep{SISR}, which implements the
(iterative) sure independence screening procedure and its extensions, as
detailed in \citet{Fan2007SISforUHD}, \citet{Fan2010sisglms},
\citet{fan2010high}. This package also provides functionality for
estimating a penalized generalized linear model or a cox regression
model for the variables selected by the screening procedure. Package
\pkg{VariableScreening} \citep{pkg:VariableScreening} offers screening
methods for independent and identically distributed (iid) data,
varying-coefficient models, and longitudinal data and includes
techniques such as sure independent ranking and screening (SIRS), which
ranks the predictors by their correlation with the rank-ordered
response, or distance correlation sure independence screening (DC-SIS),
a non-parametric extension of the correlation coefficient. Package
\pkg{MFSIS} \citep{pkg:MFSIS} provides a collection of model-free
screening techniques including SIRS, DC-SIS, the fused Kolmogorov filter
\citep{mai2015fusedkolmogorov} the projection correlation method using
knock-off features \citep{liu2020knockoff}, among others. Additional
packages implement specific procedures but their review is beyond the
scope of the current paper.

Although combining variable screening with random projection effectively
reduces the predictor set and computational costs, the variability
introduced by random sampling -- both in projections and screening
indices -- can be mitigated by averaging the results from multiple
iterations \citep{Thanei2017RPforHDR}. To address these points,
\pkg{spar} builds an ensemble of GLMs in the spirit of
\citet{Dunson2020TargRandProj} and \citet{parzer2024glms} where, in each
model of the ensemble, i) variables are first screened based on a
screening coefficient, ii) the selected variables are then projected to
a lower dimensional space, iii) a GLM is estimated using the projected
predictors. Finally, additional sparsity in the coefficients of the
original variables can be introduced through a thresholding parameter,
which together with the number of models in the ensemble can be chosen
using a validation set or via cross-validation. The final coefficients
are then obtained by averaging over the marginal models in the ensemble.
This algorithm performs sparse projected averaged regression (SPAR) for
both discrete and continuous data in the GLM framework in a
computationally efficient way. Different variants of the algorithm have
been shown to perform well in terms of prediction power on a variety of
data sets \citep[see][]{Dunson2020TargRandProj}. In particular,
\citet{parzer2024glms} show that when paired with a carefully
constructed data-driven random projection the algorithm performs
superiorly in terms of predictions and variable ranking in settings
exhibiting different degrees of sparsity in the coefficients.

A variety of screening coefficients are provided as well as several
procedures for generating random projection matrices. These procedures
can be classified into data-agnostic and data-driven, where the former,
unlike the latter, does not incorporate information from the data in the
construction of the matrices. However, the package is designed with
flexibility in mind, providing users with a versatile framework to
extend the implemented screening and projection techniques with their
own custom procedures. This is facilitated by leveraging \proglang{R}'s
\proglang{S}3 classes, making the process convenient and user-friendly.
Therefore, users can seamlessly integrate various techniques by either
writing their own procedures or leveraging the existing \proglang{R}
packages.

The package provides methods such as \texttt{plot}, \texttt{predict},
\texttt{coef}, \texttt{print}, which allow users to more easily interact
with the model output and analyze the results. The GLM framework,
especially when combined with random projections which preserve
information on the original coefficients \citep[such as the one
in][]{parzer2024glms}, facilitates interpretability of the model output,
allowing users to understand variable effects.

While \pkg{spar} offers the first implementation of the described
algorithm and, to the best of our knowledge, no other package offers the
same functionality for GLMs, few other \proglang{R} packages focus on
building ensembles where the dimensionality of the predictors is
reduced. Most notably, package \pkg{RPEnsemble} \citep{RPEnsembleR}
implements the procedure in \citet{cannings2017random} where
``carefully-selected'' random projections are used for projecting the
predictors before they are employed in a classifier such as
\(k\)-nearest neighbor, linear or quadratic discriminant analysis. On
the other hand, package \pkg{RaSEn} \citep{pkg:RaSEn} implements the
RaSE algorithm for ensemble classification and regression problems,
where random subspaces are generated and the optimal one is chosen to
train a weak learner on the basis of some criterion.

The rest of the paper is organized as follows: Section~\ref{sec-models}
provides the methodological details of the implemented algorithm. The
package is described in Section~\ref{sec-software} and
Section~\ref{sec-extensibility} exemplifies how a new screening
coefficient and a new random projection can be integrated in the
package. Section~\ref{sec-illustrations} contains two examples of
employing the package on real data sets. Finally,
Section~\ref{sec-conclusion} concludes.

\section{Methods}\label{sec-models}

The package implements a procedure for building an ensemble of GLMs
where we employ screening and random projection to the predictor matrix
pre-model estimation for the purpose of dimensionality reduction.

Throughout the section we assume to observe high-dimensional data
\(\{(\boldsymbol{x}_i,y_i)\}_{i=1}^n\), where
\(\boldsymbol{x}_i\in\mathbb{R}^p\) is a predictor vector and
\(y_i\in\mathbb{R}\) is the response, with \(p\gg n\). The predictor
vectors are collected in the rows of the predictor matrix
\(X\in \mathbb R^{n\times p}\).

\subsection{Variable screening}\label{variable-screening}

In this section we provide an overview of possible approaches to
performing variable screening in a high-dimensional setting. In
high-dimensional modeling, the goal of variable screening is to reduce
the predictor set by selecting a small subset of variables with a strong
\emph{utility} to the response variable. This initial selection enables
more efficient downstream analyses by discarding less relevant
predictors early in the modeling process, thus reducing computational
costs and potential noise accumulation stemming from irrelevant
variables \citep[see e.g.,][]{Dunson2020TargRandProj}.

Classic approaches such as sure independence screening (SIS), proposed
by \citet{Fan2007SISforUHD}, use the vector of marginal empirical
correlations
\(\hat\omega=(\omega_1,\ldots ,\omega_p)^\top\in\mathbb{R}^p,\omega_j=\text{Cor}(X_{j},y)\),
where \(y\) is the \((n\times 1)\) vector of responses and \(X_{j}\) is
the \(j\)-th column of the matrix of predictors, to screen predictors in
a linear regression setting by selecting the variable set
\(\mathcal{A}_\gamma = \{j\in [p]:|w_j|>\gamma\}\) depending on a
threshold \(\gamma>0\), where \([p]=\{1,\dots,p\}\). Under certain
technical conditions, this screening coefficient has the \emph{sure
screening property}
\(\Prob(\mathcal{A} \subset \mathcal{A}_{\gamma_n})\to 1 \text{ for } n\to \infty\),
where \(\mathcal{A}=\{j\in[p]:\beta_j\neq 0\}\) is the set of truly
active variables. Extensions to SIS include modifications for GLMs
\citep{Fan2010sisglms}, where screening is performed based on the
log-likelihood \(\ell(.)\) or the slope coefficient of the GLM
containing only \(X_j\) as a predictor:
\(\hat\omega_j=: \text{argmin}_{\beta_j\in\mathbb{R}}\text{min}_{{\beta_0}\in\mathbb{R}}\sum_{i=1}^n -\ell(\beta_j,\beta_0;y_i,x_{ij})\),
where \(x_{ij}\) is the \(j\)-th entry of the vector \(x_i\).

However, both mentioned approaches face limitations related to the
required technical conditions which can rule out practically possible
scenarios where an important variable is marginally uncorrelated to the
response due to their multicollinearity. To tackle these issues,
\citet{fan2009ultrahigh} propose to use an iterative procedure where SIS
is applied subsequently on the residuals of the model estimated in a
previous step. Additionally, in a linear regression setting,
\citet{Wang2015HOLP} propose employing the ridge estimator when the
penalty term converges to zero while \citet{cho2012high} propose using
the tilted correlation, i.e., the correlation of a tilted version of
\(X_j\) with \(y\) where the effect of other variables is reduced. For
discrete outcomes, joint feature screening \citep{SMLE2014} has been
proposed.

In order to tackle potential model misspecification, a rich stream of
literature focuses on developing semi- or non-parametric alternatives to
SIS. For linear regression, approaches include using the ranked
correlation \citep{zhu2011model}, (conditional) distance correlation
\citep{li2012feature, wang2015conditional} or quantile correlation
\citep{ma2016robust}. For GLMs, \citet{fan2011nonparametric} extend
\citet{Fan2010sisglms} by fitting a generalized additive model with
B-splines. Further extensions for discrete (or categorical) outcomes
include the fused Kolmogorov filter \citep{mai2013kolmogorov}, the mean
conditional variance, i.e., the expectation in \(X_j\) of the variance
in the response of the conditional cumulative distribution function
\(\Prob(X\leq x|Y)\) \citep{cui2015model}. \citet{ke2023sufficient}
propose a model free method where the contribution of each individual
predictor is quantified marginally and conditionally in the presence of
the control variables as well as the other candidates by
reproducing-kernel-based \(R^2\) and partial \(R^2\) statistics.

Package \pkg{spar} allows the integration of such (advanced) screening
techniques using a flexible framework, which in turn enables users to
apply various screening methods tailored to their data characteristics
in the algorithm generating the ensemble. This flexibility allows users
to evaluate different strategies, ensuring that the most effective
approach is chosen for the specific application at hand. Moreover, it
incorporates probabilistic screening strategies, which can be
particularly useful in ensembles, as they enhance the diversity of
predictors across ensemble models. Instead of relying on a fixed
threshold or number of predictors to be screened, predictors are sampled
with probabilities proportional to their screening score
\citep[see][]{Dunson2020TargRandProj, parzer2024glms}.

\subsection{Random projection tools}\label{sec-rps}

Package \pkg{spar} has been designed to allow the incorporation of
various random projection techniques, enabling users to tailor the
procedure to their specific data needs. Below, we provide background
information on random projection techniques and an overview of possible
choices for building such random projection matrices.

The random projection method relies on the Johnson-Lindenstrauss (JL)
lemma \citep{JohnsonLindenstrauss1984}, which asserts that for each set
of points in \(p\)-dimensional Euclidean space collected in the rows of
\(X\in \mathbb{R}^{n\times p}\) there exists a linear map
\(\Phi\in \mathbb{R}^{m \times p}\) such that all pairwise distances are
approximately preserved within a factor of \((1\pm\epsilon)\) for
\(m\geq m_0=\mathcal O(\epsilon^{-2}\log(n))\). Computationally, an
attractive feature of the method for high-dimensional settings is that
the bound does not depend on \(p\).

The goal is to choose a random map \(\Phi\) that satisfies the JL lemma
with high probability given that it fulfills certain technical
conditions. The literature focuses on constructing such matrices either
by sampling them from some ``appropriate'' distribution, by inducing
sparsity in the matrix and/or by employing specific fast constructs
which lead to efficient matrix-vector multiplications. It turns out that
the conditions are generally satisfied by nearly all sub-Gaussian
distributions \citep{matouvsek2008variants}. A common choice is the
standard normal distribution \(\Phi_{ij} \overset{iid}{\sim} N(0,1)\)
\citep{FRANKL1988JLSphere} or a sparser version where
\(\Phi_{ij}\overset{iid}{\sim} N(0,1/\sqrt{\psi})\) with probability
\(\psi\) and \(0\) otherwise \citep{matouvsek2008variants}. Another
computationally simpler option is the Rademacher distribution where
\(\Phi_{ij} =  \pm 1/\sqrt{\psi}\) with probability \(\psi/2\) and zero
otherwise for \(\quad 0<\psi\leq 1\), where \citet{ACHLIOPTAS2003JL}
shows results for \(\psi=1\) and \(\psi=1/3\) while
\citet{LiHastie2006VerySparseRP} recommend using \(\psi=1/\sqrt{p}\) to
obtain very sparse matrices.

Further approaches include using the Haar measure to generate random
orthogonal matrices \citep{cannings2017random} or a non-sub-Gaussian
distribution like the standard Cauchy, proposed by
\citet{LiHastie2006VerySparseRP} for preserving approximate \(\ell_1\)
distances in settings where the data is high-dimensional, non-sparse,
and heavy-tailed. Structured matrices, which allow for more efficient
multiplication, have also been proposed \citep[see
e.g.,][]{ailon2009fast, Clarkson2013LowRankApprox}.

The conventional random projections mentioned above are data-agnostic.
However, recent work has proposed incorporating information from the
data either to select the ``best'' random projection or to directly
inform the random projection procedure. For example,
\citet{cannings2017random} build an ensemble classifier where the random
projection matrix is chosen by selecting the one that minimizes the test
error of the classification problem among a set of data-agnostic random
projections.

On the other hand, \citet{parzer2024glms} propose to use a random
projection matrix for GLMs which directly incorporates information about
the relationship between the predictors and the response in the
projection matrix, rather than a projection matrix which satisfies the
JL lemma. \citet{parzer2024sparse} also provide in the linear regression
a theoretical bound on the expected gain in prediction error in using a
projection which incorporates information about the true \(\beta\)
coefficients compared to a conventional random projection. Motivated by
this result, they propose to construct a projection matrix using the
sparse embedding matrix of \citet{Clarkson2013LowRankApprox}, where the
random diagonal elements are replaced in practice by a ridge coefficient
with a minimal \(\lambda\) penalty. This method has the advantage of
approximately capturing the true beta in the span of the random
projection, i.e., it ensures that the true regression coefficients can
be recovered approximately after the projection.

Another data-driven approach to random projection for regression has
been proposed by \citet{ryder2019asymmetric}, who propose a
data-informed random projection using an asymmetric transformation of
the predictor matrix without using information of the response.

\subsection{Generalized linear models}\label{generalized-linear-models}

After we perform in each marginal model an initial screening step
followed by a projection step, we assume that the reduced and projected
set of predictors \(\boldsymbol{z}_i\) together with the response arises
from a GLM with the response having conditional density from a
(reproductive) exponential dispersion family of the form
\begin{align*}\label{eqn:y_density}
  f(y_i|\theta_i,\phi) = \exp\Bigl\{\frac{y_i\theta_i- b(\theta_i)}{a(\phi)} + c(y_i,\phi)\Bigr\},
  \quad
    g(\E[y_i|\boldsymbol{z}_i]) = \gamma_0 + \boldsymbol{z}_i^\top\boldsymbol{\gamma}=:\eta_i,
\end{align*} where \(\theta_i\) is the natural parameter, \(a(.)>0\) and
\(c(.)\) are specific real-valued functions determining different
families, \(\phi\) is a dispersion parameter, and \(b(.)\) is the
log-partition function normalizing the density to integrate to one. If
\(\phi\) is known, we obtain densities in the natural exponential family
for our responses. The responses are related to the \(m\)-dimensional
reduced and projected predictors through the conditional mean, i.e., the
conditional mean of \(y_i\) given \({\boldsymbol{z}}_i\) depends on a
linear combination of the predictors through a (invertible) link
function \(g(.)\), where \(\gamma_0\in\mathbb{R}\) is the intercept and
\(\boldsymbol{\gamma}\in\mathbb{R}^m\) is a vector of regression
coefficients for the \(m\) projected predictors.

Given that \(m\), the goal dimension of the projection need not
necessarily be small in comparison to \(n\) (we recommend using a
dimension of at most \(n/2\)), we observe that adding a small \(L_2\)
penalty in the marginal models, especially for the binomial family, can
make estimation more stable as it alleviates problems related to
separation. Therefore, we propose to employ marginal models which are
estimated by either maximizing the log-likelihood or a penalized version
thereof: \[
 \text{argmin}_{{\gamma}\in\mathbb{R}^m}\min_{\gamma_0\in\mathbb{R}}  \sum_{i=1}^n -\ell(\gamma_0, \gamma;y_i,\boldsymbol{z}_i) + \frac{\varepsilon}{2}\sum_{j=1}^m{\gamma}_j^2, \, \varepsilon \geq 0.
\] However, the framework can in principle accommodate for other penalty
functions.

\subsection{SPAR algorithm}\label{sec-algo}

We present the general algorithm for sparse projected averaged
regression (SPAR) implemented in package \pkg{spar}.

\begin{enumerate}
\def\labelenumi{\arabic{enumi}.}
\tightlist
\item
  Choose family with corresponding log-likelihood \(\ell(.)\) and link.
\item
  Standardize the \((n\times p)\) matrix of predictors \(X\) for all
  families and the vector of responses \(y\) for the Gaussian family by
  subtracting the sample column mean and dividing by the sample standard
  deviation.
\item
  Calculate screening coefficients \(\hat\omega\).
\item
  For \(k=1,\dots,M\) models:

  \begin{enumerate}
  \def\labelenumii{\arabic{enumii}.}
  \tightlist
  \item
    If \(p>2n\), screen \(2n\) predictors based on the screening
    coefficient \(\hat\omega\), which yields for model \(k\) the
    screening index set \(I_k=\{j_1^k,\dots,j_{2n}^k\}\subset[p]\); if
    probabilistic screening should be employed draw the predictors
    sequentially without replacement using an initial vector of
    probabilities \(p_j\propto |\hat\omega_j|\). Otherwise, select the
    \(2n\) variables with the highest \(|\hat\omega_j|\). If \(p < 2n\),
    perform no screening and \(I_k=\{1,\dots,p\}\).
  \item
    project screened variables to a random dimension
    \(m_k\sim \text{Unif}\{\log(p),\dots,n/2\}\) using
    \textbf{projection matrix} \(\Phi_k\) to obtain
    \(Z_k=X_{\cdot I_k}\Phi_k^\top \in \mathbb{R}^{n\times m_k}\), where
    \(X_{\cdot I_k}\) contains the columns in \(X\) having a column
    index in \(I_k\).
  \item
    it a (\(L_2\) penalized) \textbf{GLM} of \(y\) on \(Z_k\) to obtain
    estimated coefficients \(\widehat\gamma^k\in\mathbb{R}^{m_k}\) and
    \(\hat \beta_{I_k}^k=\Phi_k^\top\widehat\gamma^k\),
    \(\hat \beta_{\bar I_k}^k=0\).
  \end{enumerate}
\item
  For a given threshold \(\nu>0\), set all \(\hat\beta_j^k\) with
  \(|\hat\beta_j^k|<\nu\) to \(0\) for all \(j,k\).
\item
  Choose \(M\) and \(\nu\) via a validation set or cross-validation by
  repeating steps 1 to 4 and employing a loss function \(K(M, \nu)\) on
  the test set \begin{align*}
        (M_{\text{best}},\nu_{\text{best}}) = \text{argmin}_{M,\nu}K(M,\nu).
      \end{align*}
\item
  Combine models of the ensembles via the coefficients using
  \textbf{simple average} \(\hat \beta = \sum_{k=1}^M\hat \beta^k / M\).
\item
  Output the estimated coefficients and predictions for the chosen \(M\)
  and \(\nu\).
\end{enumerate}

\section{Software}\label{sec-software}

The package can be installed from \proglang{GitHub}

\begin{verbatim}
R> devtools::install_github("RomanParzer/SPAR")
\end{verbatim}

and loaded by

\begin{verbatim}
R> library("spar")
\end{verbatim}

In this section we rely for illustration purposes on the example data
set from the package which contains \(n=200\) observations of a
continuous response \texttt{y} and \(p=2000\) predictors \texttt{x}
which can be used as a training data set and \(n=100\) observations to
be used as a test set.

\begin{verbatim}
R> data("example_data", package = "spar")
R> str(example_data)
#> List of 7
#>  $ x     : num [1:200, 1:2000] 1.8302 -0.4251 -1.3893 -0.0947 0.4304 ...
#>  $ y     : num [1:200] -5.64 -23.63 -17.09 13.18 20.91 ...
#>  $ xtest : num [1:100, 1:2000] -0.166 -0.3729 0.0379 0.6774 0.2174 ...
#>  $ ytest : num [1:100] 10.61 -34.1 29.3 35.53 8.67 ...
#>  $ mu    : num 1
#>  $ beta  : num [1:2000] 1 -2 3 2 1 -3 2 3 1 -2 ...
#>  $ sigma2: num 83
\end{verbatim}

This data set has been simulated from a linear regression model with
\(\sigma^2=83\), an intercept \(\mu=1\) and \(\beta\) coefficients with
100 non-zero entries, where the non-zero entries are uniformly sampled
from \(\{-3,-2,-1,1,2,3\}\).

\subsection{Main functions and their
arguments}\label{main-functions-and-their-arguments}

The two main functions for fitting the SPAR algorithm are:

\begin{verbatim}
spar(x, y, family = gaussian("identity"), model = NULL, 
  rp = NULL, screencoef = NULL,
  xval = NULL, yval = NULL, nnu = 20, nus = NULL, nummods = c(20),
  measure = c("deviance", "mse", "mae", "class", "1-auc"),
  inds = NULL, RPMs = NULL, ...)
\end{verbatim}

which implements the algorithm in Section~\ref{sec-algo} without
cross-validation and returns an object of class \class{spar}, and

\begin{verbatim}
spar.cv(x, y, family = gaussian("identity"), model = NULL, 
  rp = NULL, screencoef = NULL,
  nfolds = 10, nnu = 20, nus = NULL, nummods = c(20),
  measure = c("deviance", "mse", "mae", "class", "1-auc"), ...)
\end{verbatim}

which implements the cross-validated procedure and returns an object of
class \class{spar.cv}.

The common arguments of these functions are:

\begin{itemize}
\item
  \texttt{x} an \(n \times p\) numeric matrix of predictor variables,
\item
  \texttt{y} numeric response vector of length \(n\),
\item
  \texttt{family} object from \fct{stats::family}; defaults to
  \texttt{gaussian()};
\item
  \texttt{model} an object of class \class{sparmodel} which specifies
  the model employed for each element of the ensemble.
\item
  \texttt{rp} an object of class \class{randomprojection}, defaults to
  \texttt{rp\_cw(data\ =\ TRUE)};
\item
  \texttt{screencoef} an object of class \class{screencoef}, defaults to
  \texttt{screen\_glmnet()};
\item
  \texttt{nnu} is the number of threshold values \(\nu\) which should be
  considered for thresholding; defaults to 20;
\item
  \texttt{nus} is an optional vector of \(\nu\) values to be considered
  for thresholding. If it is not provided, is defaults to a grid of
  \texttt{nnu} values. This grid is generated by including zero and
  \texttt{nnu}\(-1\) quantiles of the absolute values of the estimated
  coefficients from the marginal models, chosen to be equally spaced on
  the probability scale .
\item
  \texttt{nummods} is the number of models to be considered in the
  ensemble; defaults to 20. If a vector is provided, all combinations of
  \texttt{nus} and \texttt{nummods} are considered when choosing the
  optimal \(\nu_\text{best}\) and \(M_\text{best}\).
\item
  \texttt{measure} specifies the measure \(K(\nu, M)\) based on which
  the thresholding value \(\nu_\text{opt}\) and the number of models
  \texttt{M} should be chosen on the validation set (for
  \texttt{spar()}) or in each of the folds (in \texttt{spar.cv()}). The
  default value for \texttt{measure} is \texttt{"deviance"}, which is
  available for all families. Other options are mean squared error
  \texttt{"mse"} or mean absolute error \texttt{"mae"} (between
  responses and predicted conditional means, for all families),
  \texttt{"class"} (misclassification error) and \texttt{"1-auc"} (one
  minus area under the ROC curve) both just for binomial family.
\end{itemize}

Furthermore, \texttt{spar()} has the specific arguments:

\begin{itemize}
\item
  \texttt{xval} and \texttt{yval} which are used as validation sets for
  choosing \(\nu_\text{best}\) and \(M_\text{best}\). If not provided,
  \texttt{x} and \texttt{y} will be employed.
\item
  \texttt{inds} is an optional list of length \texttt{max(nummods)}
  containing column index-vectors corresponding to variables that should
  be kept after screening for each marginal model; dimensions need to
  fit those of the dimensions of the provided matrices in \texttt{RPM}.
\item
  \texttt{RPMs} is an optional list of length \texttt{max(nummods)}
  which contains projection matrices to be used in each marginal model.
\end{itemize}

Function \texttt{spar.cv()} has the specific argument \texttt{nfolds}
which is the number of folds to be used for cross-validation. It relies
on \texttt{spar()} as a workhorse, which is called for each fold. The
random projections for each model are held fixed throughout the
cross-validation to reduce the computational burden. This is possible by
calling \texttt{spar()} in each fold with a predefined \texttt{inds} and
\texttt{RPMs} argument, which are generated by first calling
\texttt{spar()} on the whole data set, before starting the
cross-validation procedure. However, it is possible to specify whether
the data associated with the random projection (relevant for data-driven
random projections) should be updated in each fold iteration with the
corresponding training data. This is achieved by modifying elements of
the \class{randomprojection} object, which we will exemplify in
Section~\ref{sec-extensibility}.

\subsection{Screening coefficients}\label{screening-coefficients}

The objects for creating screening coefficients are implemented as
\proglang{S}3 classes \class{screencoef}. These objects are created by
several implemented \texttt{screen\_*} functions, which take as
arguments \texttt{...} (to be saved as attributes) and \texttt{control}
(a list of controls to be used in the main function for computing the
screening coefficient).

Consider as an example function \texttt{screen\_marglik} which
implements a screening procedure based on the coefficients of univariate
GLMs:

\begin{verbatim}
R> screen_marglik
#> function(..., control = list()) {
#>     out <- list(name = name,
#>                 generate_fun = generate_fun,
#>                 control = control)
#>     attr <- list2(...)
#>     attributes(out) <- c(attributes(out), attr)
#>     if (is.null(attr(out, "type"))) {
#>       attr(out, "type") <- "prob"
#>     } else {
#>       stopifnot(
#>         "'type' must be either 'prob' or 'fixed'." =
#>           (attr(out, "type") == "prob" | attr(out, "type") == "fixed")
#>       )
#>     }
#>     class(out) <- c("screencoef")
#>     return(out)
#>   }
#> <bytecode: 0x11853e2e8>
#> <environment: 0x11852ecc0>
\end{verbatim}

Arguments related to the screening procedure can be passed through
\texttt{...}, and will be saved as attributes of the \class{screencoef}
object. More specifically, the following attributes are relevant for
function \texttt{spar()}:

\begin{itemize}
\item
  \texttt{nscreen} integer giving the number of variables to be retained
  after screening; defaults to \(2n\).
\item
  \texttt{split\_data\_prop}, double between 0 and 1 which indicates the
  proportion of the data that should be used for computing the screening
  coefficient. The remaining data will be used for estimating the
  marginal models in the SPAR algorithm; defaults to \texttt{NULL}. In
  this case the whole data will be used for estimating the screening
  coefficient and the marginal models.
\item
  \texttt{type} character -- either \texttt{"prob"} (indicating that
  probabilistic screening should be employed) or \texttt{"fixed"}
  (indicating that a fixed set of \texttt{nscreen} variables should be
  employed across the ensemble); defaults to \texttt{type\ =\ "prob"}.
\end{itemize}

The \texttt{control} argument, on the other hand, is a list containing
extra parameters to be passed to the main function computing the
screening coefficients.

The following screening coefficients are implemented in \pkg{spar}:

\begin{itemize}
\item
  \texttt{screen\_marglik()} - computes the screening coefficients by
  the coefficient of \(x_j\) for \(j =1,\dots,p\) in a univariate GLM
  using the \texttt{stats::glm()} function. \[
   \hat\omega_j=:\text{argmin}_{\beta_j\in \mathbb{R}}\text{min}_{{\beta_0}\in\mathbb{R}}\sum_{i=1}^n -\ell(\beta_0,\beta_j;y_i,x_{ij})
   \] It allows to pass a list of controls through the \texttt{control}
  argument to \texttt{stats::glm} such as weights, family, offsets.
\item
  \texttt{screen\_cor()} -- computes the screening coefficients by the
  correlation between \(y\) and \(x_j\) using the function
  \texttt{stats::cor()}. It allows to pass a list of controls through
  the \texttt{control} argument to \texttt{stats::cor}.
\item
  \texttt{screen\_glmnet()} -- computes by default the ridge coefficient
  where the penalty \(\lambda\) is very small \citep[see][ for
  clarification]{parzer2024glms}. \[
  \hat\omega=: \text{argmin}_{\beta\in \mathbb{R}^p}\text{min}_{{\beta_0}\in\mathbb{R}}\sum_{i=1}^n -\ell(\beta;y_i,x_i) + \frac{\varepsilon}{2}\sum_{j=1}^p{\beta}_j^2, \, \varepsilon > 0
  \] The function relies on \texttt{glmnet::glmnet()} and, while it
  assumes by default \(\alpha = 0\) and a small penalty, it allows to
  pass a list of controls through the \texttt{control} argument to
  \texttt{glmnet::glmnet()} such as \texttt{alpha\ =\ 1}. This screening
  coefficient is used as a default if \texttt{screencoef\ =\ NULL} in
  function call of \texttt{spar()} or \texttt{spar.cv()}.
\end{itemize}

All implemented \texttt{screen\_*} functions return an object of class
\class{screencoef} which in turn is a list with three elements:

\begin{itemize}
\item
  a character \texttt{name},
\item
  \texttt{generate\_fun()} -- an \proglang{R} function for generating
  the screening coefficient. This function should have as following
  arguments: \texttt{x} -- the matrix of standardized predictors -- and
  \texttt{y} -- the vector of (standardized in the Gaussian case)
  responses, and the argument \texttt{object}, which is a
  \class{screencoef} object itself. It returns a vector of screening
  coefficients of length \(p\).
\item
  \texttt{control}, which is the control list passed by the user in
  \texttt{screen\_*}. These controls are arguments which are needed in
  \texttt{generate\_fun()} in order to generate the desired screening
  coefficients.
\end{itemize}

For illustration purposes, consider the object created by calling
\texttt{screen\_marglik()}:

\begin{verbatim}
R> obj <- screen_marglik()
\end{verbatim}

A user-friendly \texttt{print} of the \class{screencoef} is provided:

\begin{verbatim}
R> obj
#> Name: screen_marglik 
#> Main attributes: 
#> * proportion of data used for screening: 1 
#> * number of screened variables: not provided, will default to 2n 
#> * type: probabilistic screening 
#> * screening coefficients: not yet computed from the data.
\end{verbatim}

The structure of the object is the following:

\begin{verbatim}
R> unclass(obj)
#> $name
#> [1] "screen_marglik"
#> 
#> $generate_fun
#> function(y, x, object) {
#>   if (is.null(object$control$family)) {
#>     object$control$family <- attr(object, "family")
#>   }
#>   coefs <- apply(x, 2, function(xj){
#>     glm_res <- do.call(function(...) glm(y ~ xj,  ...),
#>                        object$control)
#>     glm_res$coefficients[2]
#>   })
#>   coefs
#> }
#> <environment: namespace:spar>
#> 
#> $control
#> list()
#> 
#> attr(,"type")
#> [1] "prob"
\end{verbatim}

Function \texttt{generate\_fun()} defines the generation of the
screening coefficient. Note that it considers the controls in
\texttt{object\$control} when calling the \texttt{stats::glm()} function
(unless it is provided, the \texttt{family} argument in
\texttt{stats::glm()} will be set to the ``global'' family of the SPAR
algorithm which is assigned inside the \texttt{spar()} function an
attribute for the \class{screencoef} object).

For convenience, a constructor function
\texttt{constructor\_screencoef()} is provided, which can be used to
create new \texttt{screen\_*} functions. An example is presented in
Section~\ref{sec-extensscrcoef}.

\subsection{Random projections}\label{random-projections}

Similar to the screening procedure, the objects for creating random
projections are implemented as \proglang{S}3 classes
\class{randomprojection} and are created by functions
\texttt{rp\_*(...,\ control\ =\ list())}, which take \texttt{...} and a
list of controls \texttt{control} as arguments.

Arguments related to the random projection can be passed through
\texttt{...}, which will then be saved as attributes of the
\class{randomprojection} object. More specifically, the following
attributes are relevant in the SPAR algorithm:

\begin{itemize}
\item
  \texttt{mslow}: integer giving the minimum dimension to which the
  predictors should be projected; defaults to \(\log(p)\).
\item
  \texttt{msup}: integer giving the maximum dimension to which the
  predictors should be projected; defaults to \(n/2\).
\item
  \texttt{data}: boolean indicating whether the random projection is
  data-driven.
\end{itemize}

Note that for random projection matrices which satisfy the JL lemma,
\texttt{mslow} can be determined by employing existing results which
give a lower bound on the goal dimension in order to preserve the
distances between all pairs of points within a factor
\((1 \pm \epsilon)\). For example, \citet{ACHLIOPTAS2003JL} show
$m_0 = \log(n) (4+2\tau) / (\varepsilon^2/2 - \varepsilon^3/3)$ for probability $1-n^{-\tau}$.

The following random projections are implemented in \pkg{spar}:

\begin{itemize}
\item
  \texttt{rp\_gaussian()} -- random projection object where the
  generated matrix will have iid entries from a normal distribution
  (defaults to standard normal entries)
\item
  \texttt{rp\_sparse()} -- random projection object where the generated
  matrix will be the one in \citep{ACHLIOPTAS2003JL} with
  \texttt{psi\ =\ 1} by default.
\item
  \texttt{rp\_cw()} -- sparse embedding random projection in
  \citep{Clarkson2013LowRankApprox} for \texttt{rp\_cw(data\ =\ FALSE)}.
  Defaults to \texttt{rp\_cw(data=TRUE)}, which replaces the random
  elements on the diagonal by the ridge coefficients with a small
  penalty, as introduced in \citet{parzer2024glms}.
\end{itemize}

The \texttt{rp\_*} functions return an object of class
\class{randomprojection} which is a list with three elements:

\begin{itemize}
\item
  \texttt{name},
\item
  \texttt{generate\_fun()} function for generating the random projection
  matrix. This function should have arguments \code{rp}, which is itself
  a \class{randomprojection} object, \code{m}, the target dimension and
  a vector of indices \code{included\_vector} which indicates the column
  index of the original variables in the \code{x} matrix to be projected
  using the random projection. This is needed due to the fact that
  screening can be employed pre-projection. It can return a matrix or a
  sparse matrix of class \class{dgCMatrix} of the \pkg{Matrix} with
  \texttt{m} rows and \texttt{length(included\_vector)} columns.
\item
  \texttt{update\_data\_rp()} optional function used for data-driven
  random projections, which updates the \class{randomprojection} object
  with data information which is relevant for the random projection. The
  updating happens only once, before the start of the SPAR algorithm,
  where appropriate attributes are added to the \class{randomprojection}
  object. All relevant quantities are to be used in the
  \texttt{generate\_fun()} function. This function should have arguments
  \code{rp}, which is a \class{randomprojection} object to be updated,
  \texttt{x} -- the matrix of standardized predictors -- and \texttt{y}
  -- the vector of (standardized in the Gaussian case) responses.
  Returns a \class{randomprojection} object.
\item
  \texttt{update\_rpm\_w\_data()} optional function for updating the
  random projection matrices provided in the argument \texttt{RPMs} of
  functions \texttt{spar()} and \texttt{spar.cv()} with data-dependent
  parameters. While \texttt{update\_data\_rp} is employed only once at
  the start of the algorithm, \texttt{update\_rpm\_w\_data} specifies
  how to modify each random projection provided in \texttt{RPMs}. This
  is particularly relevant for the cross-validation procedure, which
  employs the random projection matrices generated by calling the
  \texttt{spar()} function on the whole data set before starting the
  cross-validation exercise.For example, in our implementation of the
  data-driven \texttt{rp\_cw()}, we only update \texttt{RPMs} by
  adjusting with the vector of screening coefficients computed on the
  current training data in each fold, but do not modify the random
  elements in each fold, to reduce the computational burden. Defaults to
  \texttt{NULL}. If not provided, the values of the provided
  \texttt{RPMs} do not change.
\item
  \texttt{control}, which is the control list in \texttt{rp\_*}. These
  controls are arguments needed in \texttt{generate\_fun()} in order to
  generate the desired random projection.
\end{itemize}

For illustration purposes, consider the implemented function
\texttt{rp\_gaussian()}, which generates a random projection with
entries drawn from the normal distribution. The \texttt{print} method
returns key information about the random projection procedure.

\begin{verbatim}
R> obj <- rp_gaussian()
R> obj
#> Name: rp_gaussian 
#> Main attributes: 
#> * Data-dependent: FALSE 
#> * Lower bound on goal dimension m: not provided, will default to log(p). 
#> * Upper bound on goal dimension m: not provided, will default to n/2.
\end{verbatim}

We turn to looking at the structure of the object:

\begin{verbatim}
R> unclass(obj)
#> $name
#> [1] "rp_gaussian"
#> 
#> $generate_fun
#> function(rp, m, included_vector) {
#>   p <- length(included_vector)
#>   control_rnorm <-
#>     rp$control[names(rp$control)  %in% names(formals(rnorm))]
#>   vals <- do.call(function(...)
#>     rnorm(m * p, ...), control_rnorm)
#>   RM <- matrix(vals, nrow = m, ncol = p)
#>   return(RM)
#> }
#> <environment: namespace:spar>
#> 
#> $update_data_fun
#> NULL
#> 
#> $update_rpm_w_data
#> NULL
#> 
#> $control
#> list()
#> 
#> attr(,"data")
#> [1] FALSE
\end{verbatim}

The \texttt{generate\_fun()} function returns a matrix with \texttt{m}
rows and \texttt{length(included\_vector)} columns. Note that
\texttt{included\_vector} gives the indices of the variables which have
been selected by the screening procedure. In this case, where the random
projection does not use any data information, we are only interested in
the length of this vector.

The functions \texttt{update\_data\_fun()} and
\texttt{update\_rpm\_w\_data()} are \texttt{NULL} as this conventional
random projection is data-agnostic.

\subsection{Marginal models}\label{marginal-models}

The package provides a class \class{sparmodel} for the marginal model to
be fitted for each element of the ensemble. The framework currently
assumes that the linear predictor is a linear combination of the
projected variables.

Similar to the objects for random projection and screening coefficients,
the functions which create these objects have arguments \texttt{...} (to
be saved as attributes) and \texttt{control} (to be used in the main
function for building the model).

The two functions implemented are \texttt{spar\_glmnet()}, which allows
regularized GLMs as marginal models using function
\texttt{glmnet::glmnet()} (where the default is to estimate a ridge
regression with the small penalty value), and \texttt{spar\_glm()} which
estimates unregularized GLMs using \texttt{stats::glm.fit()}.

An object of class \class{sparmodel} is a list with elements:

\begin{itemize}
\item
  \texttt{name},
\item
  \texttt{model\_fun()} -- a function which takes \texttt{y} (the vector
  of standardized responses), \texttt{z} (the matrix of reduced
  predictors) and a further argument which is the object of class
  \class{sparmodel} itself.
\item
  \texttt{update\_model()} -- an optional function which can add further
  attributes to the \class{sparmodel} object which is called at the
  beginning of the SPAR algorithm. In the case of
  \texttt{spar\_glmnet()} this function manipulates the \class{family}
  object in a way which is convenient for \texttt{glmnet::glmnet()}.
  \footnote{In the case of families Gaussian, binomial and Poisson with 
    canonical link, the family object is replaced by a string containing the 
    name of the family. This leads to [glmnet]{.pkg} using the faster specialized 
    algorithms rather than the general algorithm implemented for all 
    [family]{.class} objects.}
\end{itemize}

The default is to use \texttt{spar\_glm()} for Gaussian family with
identity link and \texttt{spar\_glmnet()} for the other families.

\subsection{Methods}\label{methods}

Methods \texttt{print}, \texttt{plot}, \texttt{coef}, \texttt{predict}
are available for both \class{spar} and \class{spar.cv} classes.

\subsubsection{print}\label{print}

The \texttt{print} method returns information on \(\nu_\text{best}\)
\(M_\text{best}\), the number of active predictors (i.e., predictors
which have at least a nonzero coefficient across the marginal models)
and a five-point summary of the non-zero coefficients.

\begin{verbatim}
R> set.seed(12)
R> spar_res <- spar(example_data$x, example_data$y,
+                 xval = example_data$xtest,
+                 yval = example_data$ytest,
+                 nummods = c(5,10,15,20,25,30))
R> spar_cv <- spar.cv(example_data$x, example_data$y,
+                   nummods = c(5,10,15,20,25,30))
\end{verbatim}

\begin{verbatim}
R> spar_res
#> spar object:
#> Smallest Validation Measure reached for nummod=30,
#>               nu=1.72e-02 leading to 1002 / 2000 active predictors.
#> Summary of those non-zero coefficients:
#>     Min.  1st Qu.   Median     Mean  3rd Qu.     Max. 
#> -1.19688 -0.07144  0.02134  0.02814  0.12008  0.99405
\end{verbatim}

For \class{spar.cv} it also provides the same information for the
\((\nu, M)\) combination chosen by the one-standard error rule.

\begin{verbatim}
R> spar_cv
#> spar.cv object:
#> Smallest CV-Meas 2230.9 reached for nummod=30,
#>               nu=0.00e+00 leading to 1825 / 2000 active predictors.
#> Summary of those non-zero coefficients:
#>      Min.   1st Qu.    Median      Mean   3rd Qu.      Max. 
#> -0.825363 -0.041159  0.004422  0.018693  0.072914  0.953501 
#> 
#> Sparsest coefficient within one standard error of best CV-Meas
#>               reached for nummod=5, nu=5.40e-03 
#> leading to 1003 / 2000 active
#>               predictors with CV-Meas 2748.4.
#> Summary of those non-zero coefficients:
#>     Min.  1st Qu.   Median     Mean  3rd Qu.     Max. 
#> -1.01802 -0.10460  0.04862  0.03436  0.17091  1.32992
\end{verbatim}

\subsubsection{coef}\label{coef}

Method \texttt{coef} takes as inputs a \class{spar} or \class{spar.cv}
object, together with further arguments:

\begin{itemize}
\item
  \texttt{nummod} -- number of models used to compute the averaged
  coefficients; value of \texttt{nummod} with minimal \texttt{measure}
  is used if not provided.
\item
  \texttt{nu} -- threshold level used to compute the averaged
  coefficients; value with minimal \texttt{measure} is used if not
  provided.
\end{itemize}

\begin{verbatim}
R> str(coef(spar_res))
#> List of 4
#>  $ intercept: num 2.86
#>  $ beta     : num [1:2000] 0 0 0.746 0 0 ...
#>  $ nummod   : num 30
#>  $ nu       : num 0.0172
\end{verbatim}

It returns a list with the intercept, vector of \texttt{beta}
coefficients and the \texttt{nummod} and \texttt{nu} employed in the
calculation.

Additionally for \class{spar.cv}, the \texttt{coef} method also has
argument \texttt{opt\_par} which is one of \texttt{c("1se","best")} and
chooses whether to select the best pair of \texttt{nus} and
\texttt{nummods} according to cross-validation \texttt{measure}, or the
solution yielding sparsest vector of coefficients within one standard
deviation of that optimal cross-validation \texttt{measure}. This
argument is ignored when \texttt{nummod} and \texttt{nu} are given.

\subsubsection{predict}\label{predict}

Functionality for computing predictions is provided through the method
\texttt{predict} which takes a \class{spar} or \class{spar.cv} object,
together with

\begin{itemize}
\item
  \texttt{xnew} -- matrix of new predictor variables; must have same
  number of columns as \texttt{x}.
\item
  \texttt{type} -- the type of required predictions; either on
  \texttt{"response"} level (default) or on \texttt{"link"} level
\item
  \texttt{avg\_type} -- type of averaging used across the marginal
  models; either on \texttt{"link"} (default) or on \texttt{"response"}
  level
\item
  \texttt{nummod} -- number of models used to compute the averaged
  coefficients; value of \texttt{nummod} with minimal \texttt{measure}
  is used if not provided.
\item
  \texttt{nu} -- threshold level used to compute the averaged
  coefficients; value with minimal \texttt{measure} is used if not
  provided.
\item
  \texttt{coef} -- optional vector of coefficients to be used directly
  in the prediction.
\end{itemize}

Additionally, for class \class{spar.cv}, argument \texttt{opt\_par} is
available and used in the computation of the coefficients to be used for
prediction (see above description of method \texttt{coef}).

\subsubsection{plot}\label{plot}

Plotting functionality is provided through the \texttt{plot} method,
which takes a \class{spar} or \class{spar.cv} object, together with
further arguments:

\begin{itemize}
\item
  \texttt{plot\_type} -- one of:

  \begin{itemize}
  \item
    \texttt{"Val\_Measure"} plots the (cross-)validation
    \texttt{measure} for either a grid of \texttt{nu} values for a fixed
    number of models \texttt{nummod} or viceversa.
  \item
    \texttt{"Val\_numAct"} plots the number of active variables for
    either a grid of \texttt{nu} values for a fixed number of models
    \texttt{nummod} or viceversa.
  \item
    \texttt{"res-vs-fitted"} produces a residuals-vs-fitted plot. The
    residuals are computed as \(y- \widehat y\), where \(\widehat y\) is
    the prediction computed on response level.
  \item
    \texttt{"coefs"} produces a plot of the value of the standardized
    coefficients for each predictor in each marginal model (before
    thresholding). For each predictor, the values of the coefficients
    are sorted from largest to smallest across the marginal models and
    then represented in the plot.
  \end{itemize}
\item
  \texttt{plot\_along} -- one of \texttt{c("nu","nummod")}; for
  \texttt{plot\_type\ =\ "Val\_Measure"} or
  \texttt{plot\_type\ =\ "Val\_numAct"} it indicates whether the values
  of the cross-validation measure or number of active variables,
  respectively, should be shown for a grid of \(\nu\) values while
  keeping the number of models \texttt{nummod} fixed or viceversa. This
  argument is ignored when \texttt{plot\_type\ =\ "res-vs-fitted"} or
  \texttt{plot\_type\ =\ "coefs"}.
\item
  \texttt{nummod} -- fixed value for number of models when
  \texttt{plot\_along\ =\ "nu"} for
  \texttt{plot\_type\ =\ "Val\_Measure"} or \texttt{"Val\_numAct"}; if
  \texttt{plot\_type\ =\ "res-vs-fitted"}, it is used in the
  \texttt{predict} method, as described above.
\item
  \texttt{nu} -- fixed value for \(\nu\) when
  \texttt{plot\_along\ =\ "nummod"} for
  \texttt{plot\_type\ =\ "Val\_Measure"} or \texttt{"Val\_numAct"}; if
  \texttt{plot\_type\ =\ "res-vs-fitted"}, it is used in the
  \texttt{predict} method, as described above.
\item
  \texttt{xfit} -- if \texttt{plot\_type\ =\ "res-vs-fitted"}, it is the
  matrix of predictors used in computing the fitted values. This
  argument must be provided for the plot of residuals and fitted values,
  as the \class{spar} or \class{spar.cv} objects do not store the
  original data.
\item
  \texttt{yfit} -- if \texttt{plot\_type\ =\ "res-vs-fitted"}, vector of
  responses used in computing the residuals. This argument must be
  provided for the plot of residuals and fitted values, as the
  \class{spar} or \class{spar.cv} objects do not store the original
  data.
\item
  \texttt{prange} -- optional vector of length 2 in case
  \texttt{plot\_type\ =\ "coefs"} which gives the limits of the
  predictors' plot range; defaults to \texttt{c(1,\ p)}.
\item
  \texttt{coef\_order} -- optional index vector of length \(p\) in case
  \texttt{plot\_type\ =\ "coefs"} to give the order of the predictors;
  defaults to \texttt{1\ :\ p}.
\end{itemize}

For class \class{spar.cv} there is the extra argument
\texttt{opt\_par\ =\ c("best",\ "1se")} which, for
\texttt{plot\_type\ =\ "res-vs-fitted"} indicates whether the
predictions should be based on coefficients using the best \((\nu, M)\)
combination or on the combination which delivers the sparsest \(\beta\)
having validation measure within one standard deviation from the minimum

The \texttt{plot} methods return objects of class \class{ggplot}
\citep{ggplotR}.

\section{Extensibility}\label{sec-extensibility}

\subsection{Screening coefficients}\label{sec-extensscrcoef}

We exemplify how new screening coefficients implemented in package
\pkg{VariableScreening} can easily be used in the framework of
\pkg{spar}.

We start by defining the function for generating the screening
coefficients using the \texttt{screenIID()} function in
\pkg{VariableScreening}.

\begin{verbatim}
R> generate_scr_sirs <- function(y, x, object) {
+    res_screen <- do.call(function(...) 
+      VariableScreening::screenIID(x, y, ...), 
+      object$control)
+    coefs <- res_screen$measurement
+    coefs
+  }
\end{verbatim}

Note that \texttt{screenIID()} also takes method as an argument. To
allow for flexibility, we do not fix the method in
\texttt{generate\_scr\_sirs()} but rather allow the user to pass a
method through the \texttt{control} argument in the \texttt{screen\_*}
function. This function is created using the helper
\texttt{constructor\_screencoef()}:

\begin{verbatim}
R> screen_sirs <- constructor_screencoef(
+    "screen_sirs", 
+    generate_fun = generate_scr_sirs)
\end{verbatim}

We now call the \texttt{spar()} function with the newly created
screening procedure. We consider the method SIRS of
\citet{zhu2011model}, which ranks the predictors by their correlation
with the rank-ordered response and we do not perform probabilistic
variable screening but employ the top \(2n\) variables in each marginal
model.

\begin{verbatim}
R> set.seed(123)      
R> spar_example <- spar(example_data$x, example_data$y,
+    screencoef = screen_sirs(type = "fixed",
+        control=list(method = "SIRS")),
+    measure = "mse")
R> spar_example
#> spar object:
#> Smallest Validation Measure reached for nummod=20,
#>               nu=1.75e-03 leading to 396 / 2000 active predictors.
#> Summary of those non-zero coefficients:
#>       Min.    1st Qu.     Median       Mean    3rd Qu.       Max. 
#> -0.6918230 -0.0928444  0.0009116  0.0943062  0.1777889  2.0089709
\end{verbatim}

\subsection{Random projections}\label{random-projections-1}

We exemplify how new random projections can be implemented in the
framework of \pkg{spar}.

We implement the random projection of \citet{cannings2017random}, who
propose using the Haar measure for generating the random projections.
They simulate matrices from the Haar measure by independently drawing
each entry of a matrix \(Q\) from a standard normal distribution, and
then to take the projection matrix to be the transpose of the matrix of
left singular vectors in the singular value decomposition of \(Q\).
Moreover, they suggest using ``good'' random projections, in the sense
that they deliver the best out-of-sample prediction. The proposed
approach employs \(B_1\) models in an ensemble of classifiers and for
each model \(k\), \(B_2\) data independent random projections are
generated and the one with the lowest error on a test set is the one
chosen to project the variables in model \(k\).

We can implement such a random projection in \pkg{spar} by the following
building block:

\begin{verbatim}
R> update_data_cannings <- function(rp, x, y) {
+    attr(rp, "dataset") <- list(x = x, y = y)
+    rp
+  }
\end{verbatim}

This is the function which adds data information to the random
projection object. Here, the whole data can be added as information for
the \(M\) random projection (alternatively, one could only pass
sufficient statistics for computing the desired measures).

While the \(B_2\) random projections are data-agnostic, the
\texttt{generate\_fun()} element of the random projection will need the
data information in order to evaluate which method performs best in
terms of an error measure. We will, in the following, define the
function for the generation of the random projection matrix to be used
in a model \(k\).

This helper simulates \(m\times p\) matrices from the Haar measure:

\begin{verbatim}
R> simulate_haar <- function(m, p) {
+    R0 <- matrix(1/sqrt(p) * rnorm(p * m), nrow = p, ncol = m)
+    RM <- qr.Q(qr(R0))[, seq_len(m)]
+    t(RM)
+  }
\end{verbatim}

The function that generates the random projection matrix for model \(k\)
uses 25\% of the data as a test set. After estimating a ridge regression
with a minimal penalty on the training data (this is the marginal model
to be employed in the SPAR algorithm), it chooses the best among \(B_2\)
random projections in terms of minimizing misclassification error for
the binomial family and MSE for all other families on the test set:

\begin{verbatim}
R> generate_cannings <- function(rp, m, included_vector) {
+    x <- attr(rp, "dataset")$x[, included_vector]
+    y <- attr(rp, "dataset")$y
+    n <- nrow(x);  p <- ncol(x)
+    B2 <-  ifelse(is.null(rp$control$B2), 50, rp$control$B2)
+    id_test <- sample(n, size = n %/% 4)
+    xtrain <- x[-id_test, ];  xtest <- x[id_test,]
+    ytrain <- y[-id_test];  ytest <- y[id_test]
+    if (is.null(rp$control$family)) rp$control$family <- attr(rp, "family")
+    if (is.null(rp$control$alpha)) rp$control$alpha <- 1
+    control_glmnet <-
+      rp$control[names(rp$control) %in% names(formals(glmnet::glmnet))]
+    error_all <- lapply(seq_len(B2), FUN = function(s){
+      RM <- simulate_haar(m, p)
+      xrp <- tcrossprod(xtrain, RM)
+      mod <- do.call(function(...) 
+        glmnet::glmnet(x =  xrp, y = ytrain, ...), control_glmnet)
+      coefs <- coef(mod, s = min(mod$lambda))
+      eta_test <- (cbind(1, tcrossprod(xtest, RM)) %*% coefs)
+      pred <- rp$control$family$linkinv(as.vector(eta_test))
+      out <-  ifelse(rp$control$family$family == "binomial",
+                     mean(((pred > 0.5) + 0) != ytest), 
+                     mean((pred - ytest)^2))
+      list(RM, out)
+   })
+   id_best <- which.min(sapply(error_all, "[[", 2))
+   RM <- error_all[[id_best]][[1]]
+   return(RM)
+  }
\end{verbatim}

In the cross-validation procedure, we do not generate new matrices for
each step to keep computational costs low, so we do not specify a
function \texttt{update\_rpm\_w\_data()}.

Putting it all together, we get:

\begin{verbatim}
R> rp_cannings <- constructor_randomprojection(
+    "rp_cannings",
+    generate_fun = generate_cannings,
+    update_data_fun = update_data_cannings
+  )
\end{verbatim}

We can now estimate SPAR for a binomial model, where we transform the
response to a binary variable.

\begin{verbatim}
R> ystar <- (example_data$y > 0) + 0
R> ystarval <- (example_data$ytest > 0) + 0
\end{verbatim}

We use \(50\) models (which is in line to recommendations for \(B_1\) in
\citet{cannings2017random}), and no screening procedure. Note that, if
screening should not be performed, \texttt{nscreen} can be set to
\texttt{p} in the \texttt{screen\_*} function. Moreover, we do not
perform any thresholding.

\begin{verbatim}
R>   set.seed(12345) 
R> spar_example_1 <- spar(x = example_data$x, y = ystar,
+    family = binomial(),
+    screencoef = screen_marglik(nscreen = ncol(example_data$x)),
+    rp = rp_cannings(control = list(B2 = 50, lambda.min.ratio = 0.01)),
+    nus = 0, nummods = 50, 
+    xval = example_data$xtest, yval = ystarval,
+    measure = "class"
+  )
\end{verbatim}

Using the data-driven \texttt{rp\_cw()}:

\begin{verbatim}
R>   set.seed(12345)  
R> spar_example_2 <- spar(x = example_data$x, y = ystar,
+    family = binomial(),
+    screencoef = screen_marglik(nscreen = ncol(example_data$x)),
+    rp = rp_cw(data = TRUE),
+    nus = 0, nummods = 50, 
+    xval = example_data$xtest, yval = ystarval,
+    measure = "class"
+  )
\end{verbatim}

We can now compare the two approaches by looking at the minimum measure
\texttt{Meas}\\
achieved on the validation set:

\begin{verbatim}
R> spar_example_1$val_res[which.min(spar_example_1$val_res$Meas),]
#>   nnu nu nummod numAct Meas
#> 1   1  0     50   2000 0.17
R> spar_example_2$val_res[which.min(spar_example_2$val_res$Meas),]
#>   nnu nu nummod numAct Meas
#> 1   1  0     50   2000 0.17
\end{verbatim}

\section{Illustrations}\label{sec-illustrations}

\subsection{Face image data}\label{face-image-data}

We illustrate the functionality of \pkg{spar} on the Isomap data set
containing \(n = 698\) black and white face images of size
\(p = 64 \times 64 = 4096\) together with the faces' horizontal looking
direction angle as the response variable.\footnote{
The Isomap face data can be found online on https://web.archive.org/web/20160913051505/http://isomap.
stanford.edu/datasets.html.}

\begin{verbatim}
R> url1 <- "https://web.archive.org/web/20150922051706/"
R> url2 <- "http://isomap.stanford.edu/face_data.mat.Z"
R> download.file(paste0(url1, url2),
+              file.path("face_data.mat.Z"))
R> system('uncompress face_data.mat.Z')
\end{verbatim}

The \texttt{.mat} file format can be read using \pkg{R.matlab}
\citep{pkg:rmatlab}:

\begin{verbatim}
R> library("R.matlab")
R> facedata <- readMat(file.path("face_data.mat"))
R> x <- t(facedata$images)
R> y <- facedata$poses[1,]
\end{verbatim}

We can visualize one observation in this data set in the left panel
Figure \ref{fig:faces_predictions} (code for reproducing the figure is
provided in the supplementary materials).

This data set has the issue of many columns being almost constant, which
can make estimation unstable. Given that \texttt{spar()} and
\texttt{spar.cv()} ignore constant columns, we can alleviate this issue
by setting all columns which have a low standard deviation to zero.

\begin{verbatim}
R> x[, apply(x, 2, sd) < 0.01] <- 0
\end{verbatim}

We split the data into training vs test sample

\begin{verbatim}
R> set.seed(123)
R> ntot <- length(y); ntest <- ntot * 0.25
R> testind <- sample(ntot, ntest, replace = FALSE)
R> xtrain <- as.matrix(x[-testind, ]); ytrain <- y[-testind]
R> xtest <- as.matrix(x[testind, ]); ytest <- y[testind]
\end{verbatim}

We now estimate on the training data the SPAR algorithm with
cross-validation. We employ the data-driven random projection proposed
in \citet{parzer2024sparse} with screening based on the ridge
coefficients. To ensure convergence of the \pkg{glmnet} algorithm, we
set the \texttt{lambda.min.ratio} parameter to \(0.001\). Moreover, each
marginal model is a ridge linear regression, with a small penalty (the
minimal penalty produced by \texttt{glmnet::glmnet()}).

\begin{verbatim}
R> library("spar")
R> set.seed(123)
R> control_glmnet <- list(lambda.min.ratio = 0.001)
R> spar_faces <- spar.cv(
+  xtrain, ytrain,
+  model = spar_glmnet(control = control_glmnet),
+  screencoef = screen_glmnet(control = control_glmnet),
+  rp = rp_cw(data = TRUE, control = control_glmnet),
+  nummods = c(5, 10, 20, 50),
+  measure = "mse")
\end{verbatim}

The \texttt{print} method returns:

\begin{verbatim}
R> spar_faces
#> spar.cv object:
#> Smallest CV-Meas 19.4 reached for nummod=10,
#>               nu=0.00e+00 leading to 3056 / 4096 active predictors.
#> Summary of those non-zero coefficients:
#>      Min.   1st Qu.    Median      Mean   3rd Qu.      Max. 
#> -6.091869 -0.100810  0.001171  0.032695  0.117214  8.749217 
#> 
#> Sparsest coefficient within one standard error of best CV-Meas
#>               reached for nummod=5, nu=1.85e-03 
#> leading to 1614 / 4096 active
#>               predictors with CV-Meas 25.2.
#> Summary of those non-zero coefficients:
#>     Min.  1st Qu.   Median     Mean  3rd Qu.     Max. 
#> -8.36989 -0.25699  0.06919  0.05933  0.33964  9.90484
\end{verbatim}

\begin{figure}[t!]
\centering
\includegraphics[width=0.8\textwidth]{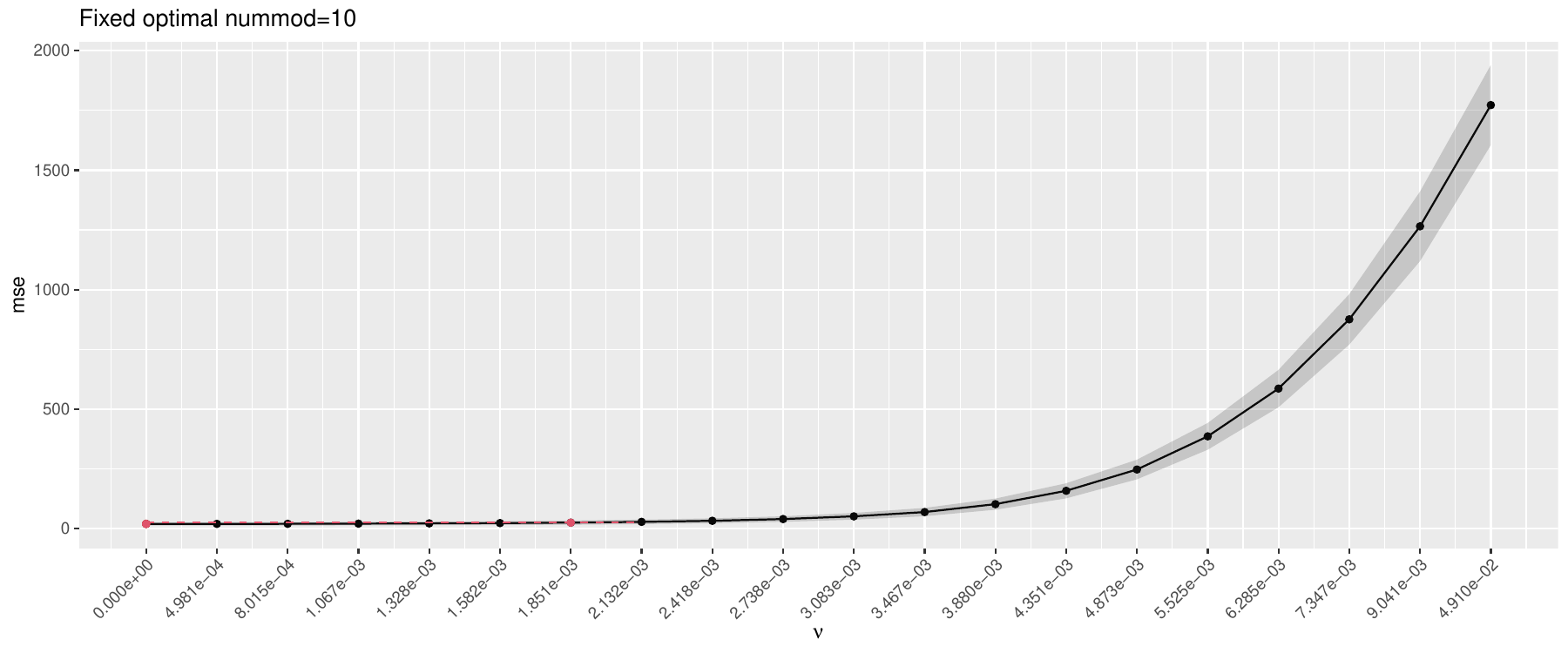}
\caption{Plot of mean squared error over a grid of threshold values $\nu$ for fixed number of optimal models $M=20$ for the \emph{Isomap faces} data set.
The red points correspond to the threshold which achieves the lowest
cross-validation measure and the one with the largest cross-validation 
measure within one standard deviation away from minimum. Confidence bands represent one standard deviation in the measures across the number of folds. \label{fig:facesplot_valmeasure}}
\end{figure}

The \texttt{plot} method for \class{spar.cv} objects displays by default
the measure employed in the cross-validation (in this case MSE) for a
grid of \(\nu\) values, where the number of models is fixed to the value
found to perform best in cross-validation exercise (see Figure
\ref{fig:facesplot_valmeasure}).

\begin{verbatim}
R> plot(spar_faces)
\end{verbatim}

We observe that no thresholding delivers the lowest MSE but that the
differences in MSE compared to the value of \(\nu=0.00158\) chosen by
1-standard-error rule are minimal.

The coefficients of the different variables (in this example pixels)
obtained by averaging over the coefficients the marginal models (for
optimal \(\nu\) and number of models) are given by:

\begin{verbatim}
R> face_coef <- coef(spar_faces, opt_par = "best")
R> str(face_coef)
#> List of 4
#>  $ intercept: num -21.8
#>  $ beta     : num [1:4096] 0 0.0392 0.52 0 -0.1647 ...
#>  $ nummod   : num 10
#>  $ nu       : num 0
\end{verbatim}

For a sparser solution we can compute the coefficients using
\texttt{opt\_par\ =\ "1se"} which leads to more sparsity and a lower
number of models.

\begin{verbatim}
R> face_coef_1se <- coef(spar_faces, opt_par = "1se")
R> str(face_coef_1se)
#> List of 4
#>  $ intercept: num -21.9
#>  $ beta     : num [1:4096] 0 0 0 0 -0.329 ...
#>  $ nummod   : num 5
#>  $ nu       : num 0.00185
\end{verbatim}

\begin{figure}[t!]
\centering
\includegraphics[width=0.8\textwidth]{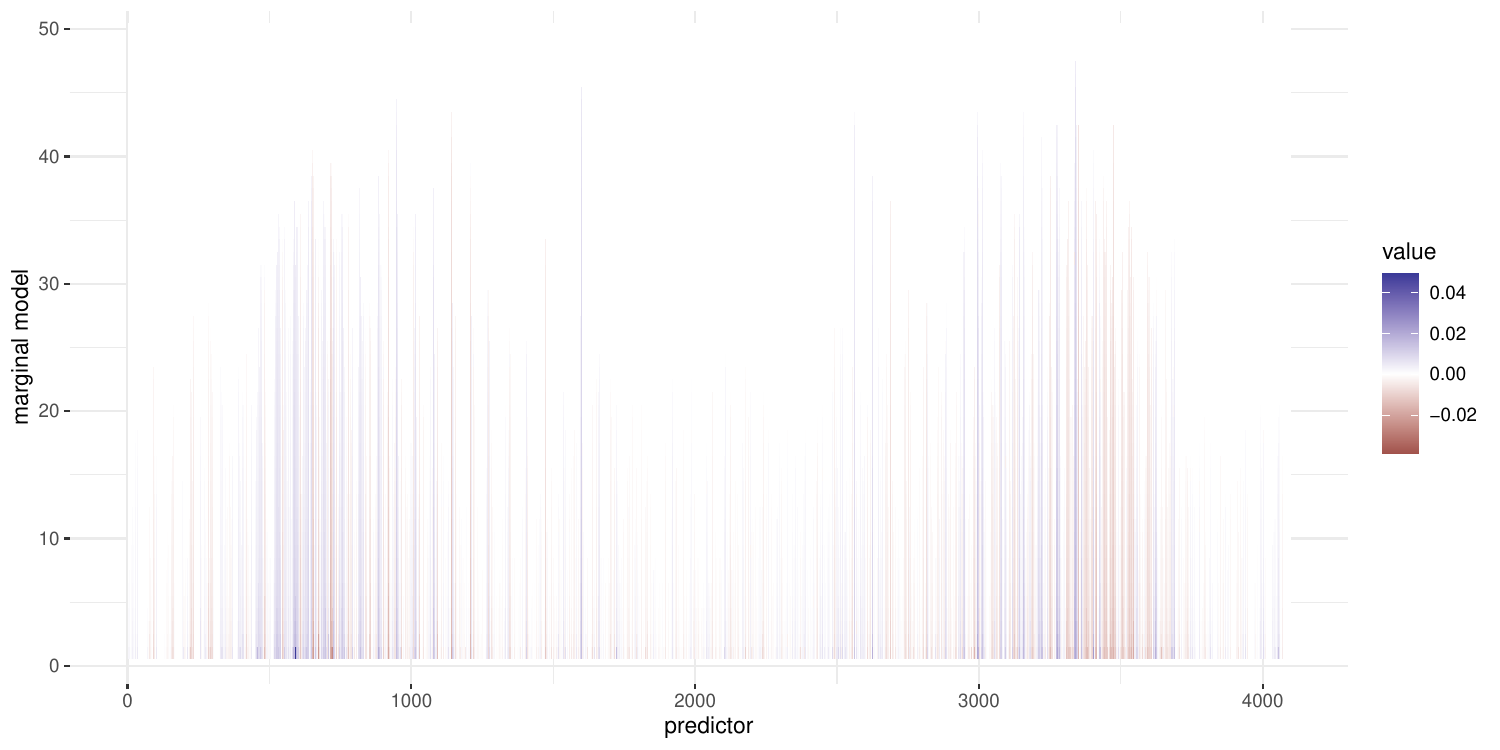}
\caption{Coefficient plot for all variables and all $M=50$ models in the SPAR ensemble for the \emph{Isomap faces} data set. The coefficients are standardized, before thresholding.
\label{fig:faces_coefs}}
\end{figure}

The standardized coefficients from each of \texttt{max(nummods)} models
(before thresholding) can be plotted by setting
\texttt{plot\_type\ =\ "coefs"} (see Figure \ref{fig:faces_coefs}).

\begin{verbatim}
R> plot(spar_faces, plot_type = "coefs")
\end{verbatim}

We observe that pixels close to each other have more correlation than
pixels further apart.

The \texttt{predict()} function can be applied to the \class{spar.cv}
object. We will employ the sparser solution chosen by the
\texttt{opt\_par\ =\ "1se"} rule:

\begin{verbatim}
R> ynew <- predict(spar_faces, xnew = xtest, coef = face_coef_1se)
\end{verbatim}

For this data set, one can visualize the effect of each pixel
\(\hat\beta^\text{1se}_j x^\text{new}_{i,j}\) in predicting the face
orientation in a given image. The contribution of each pixel in
predicting the orientation of the face in the left panel of Figure
\ref{fig:faces_predictions} can be visualized in the right panel of
Figure \ref{fig:faces_predictions}.

\begin{figure}[t!]
\centering
\includegraphics[width=0.4\textwidth]{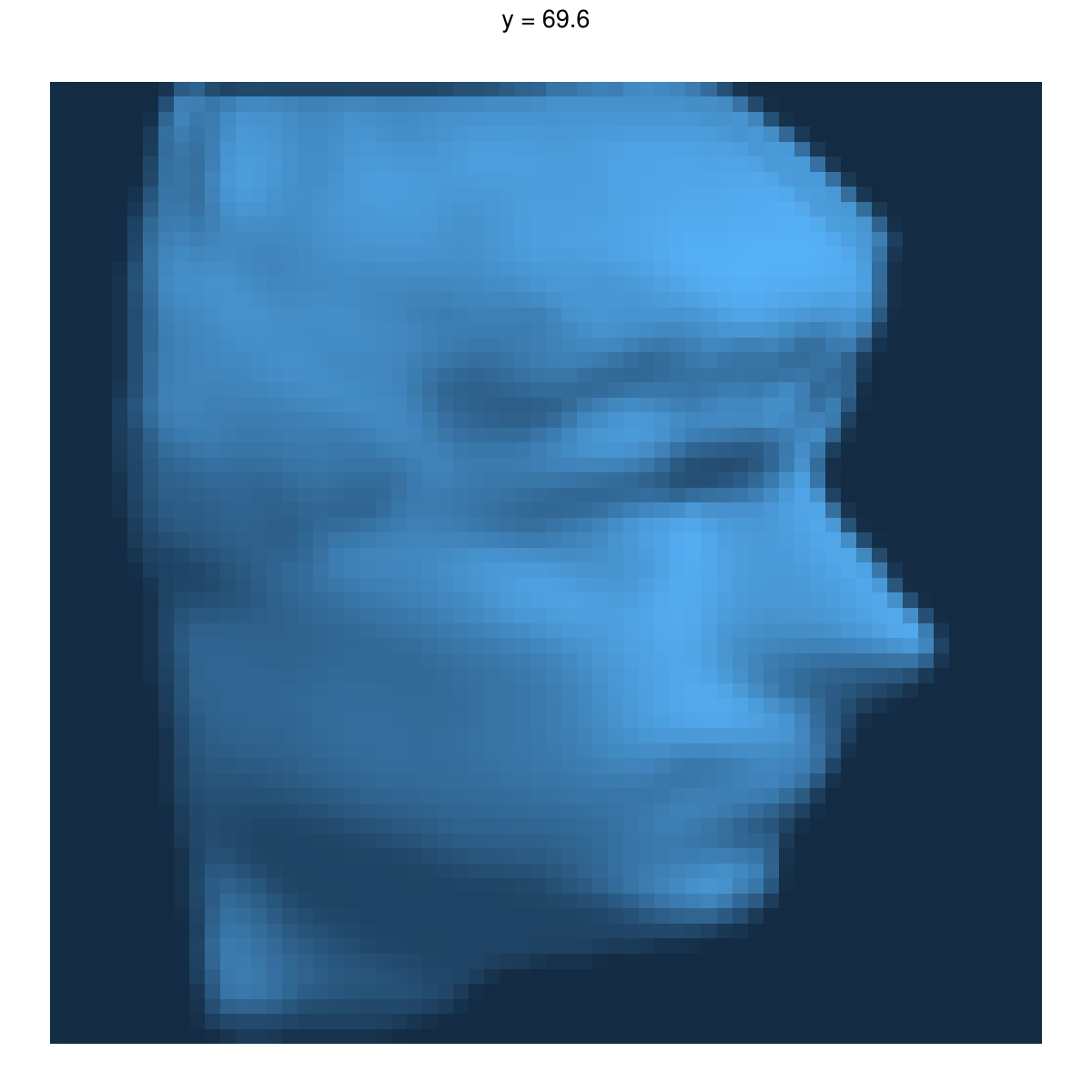}
\includegraphics[width=0.4\textwidth]{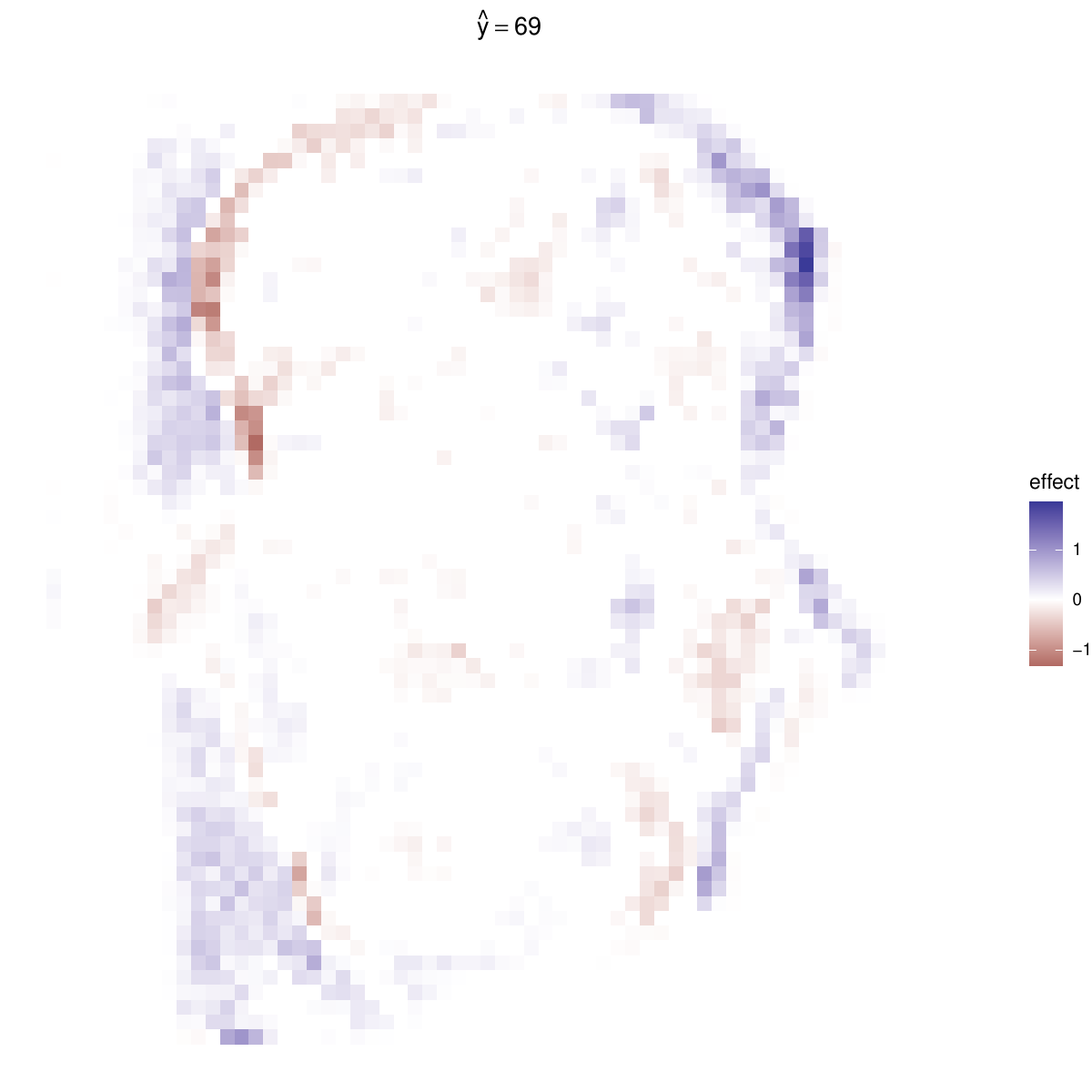}
\caption{Left: Image corresponding to one observation in the \emph{Isomap faces} data set. Right: The effect of each pixel $\hat\beta^\text{1se}_j x^\text{new}_{i,j}$ in predicting the face orientation in left panel. \label{fig:faces_predictions}}
\end{figure}

\subsection{Darwin data set}\label{darwin-data-set}

The Darwin data set \citep{CILIA2022darwin} contains a binary response
for Alzheimer's disease (AD) together with extracted features from 25
handwriting tests (18 features per task) for 89 AD patients and 85
healthy people
(\(n=174\)).\footnote{The data set can be downloaded from  https://archive.ics.uci.edu/dataset/732/darwin}

\begin{verbatim}
R> download.file("https://archive.ics.uci.edu/static/public/732/darwin.zip",
+    "darwin.zip")
\end{verbatim}

\begin{verbatim}
R> darwin_tmp <- read.csv(unzip("darwin.zip",  "data.csv"), 
+   stringsAsFactors = TRUE)
\end{verbatim}

Before proceeding with the analysis, the data is screened for
multivariate outliers using the DDC algorithm in package \pkg{cellWise}
\citep{rcellwise}.

\begin{verbatim}
R> darwin_orig <- list(
+  x = darwin_tmp[, !(colnames(darwin_tmp) %in% c("ID", "class"))],
+  y = as.numeric(darwin_tmp$class) - 1)
R> tmp <- cellWise::DDC(
+  as.matrix(darwin_orig$x),
+  list(returnBigXimp = TRUE, 
+       tolProb = 0.999,
+       silent = TRUE))
#>  
#>  The final data set we will analyze has 174 rows and 446 columns.
#> 
R> darwin <- list(x = tmp$Ximp, y = darwin_orig$y)
\end{verbatim}

The structure of the data is:

\begin{verbatim}
R> str(darwin)
#> List of 2
#>  $ x: num [1:174, 1:450] 5160 3721 2600 2130 2310 ...
#>   ..- attr(*, "dimnames")=List of 2
#>   .. ..$ : NULL
#>   .. ..$ : chr [1:450] "air_time1" "disp_index1" "gmrt_in_air1" "gmrt_on_paper1" ...
#>  $ y: num [1:174] 1 1 1 1 1 1 1 1 1 1 ...
\end{verbatim}

\begin{figure}[t!]
\centering
\includegraphics[width=0.8\textwidth]{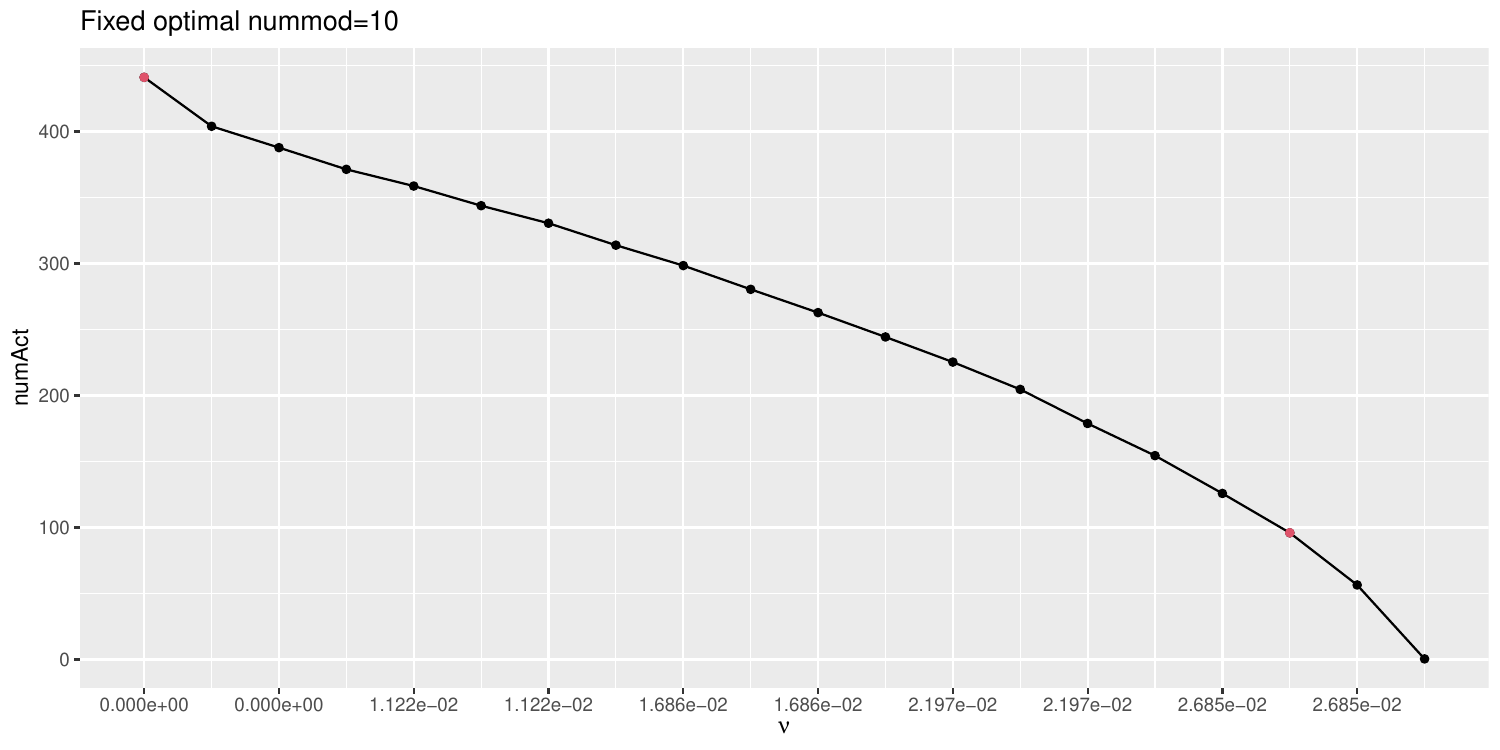}
\caption{Average number of active variables for the grid of thresholding values 
$\nu$ and $M=10$ models for the \emph{Darwin} data set. The red points 
correspond to the average number of active variables for the model with the 
lowest cross-validation measures and to the one chosen by the 1-standard-error 
rule. \label{fig:darwin_activevars}}
\end{figure}

We estimate the SPAR algorithm with the screening and random projection
introduced in \citet{parzer2024glms} for binomial family and logit link,
using \(1-\)area under the ROC curve as the cross-validation measure:

\begin{verbatim}
R> spar_darwin <- spar.cv(darwin$x, darwin$y,
+                       family = binomial(logit),
+                       nummods = c(5, 10, 20, 50),
+                       measure = "1-auc")
\end{verbatim}

We can look at the average number of active variables for a grid of
\(\nu\) where the number of models is fixed to the value found to
perform best in cross-validation exercise (see Figure
\ref{fig:darwin_activevars}).

\begin{verbatim}
R> plot(spar_darwin, plot_type = "Val_numAct")
\end{verbatim}

We observe again that no thresholding achieves the best measure and
translates to almost all variables being active (some variables can be
inactive at \(\nu=0\) as they may never be screened). The
1-standard-error rule would however indicate that more sparsity can be
introduced without too much increase in the cross-validation measure.

Finally, coefficients of the predictors across the maximum number of
considered marginal models (in this case \(M=50\)) can be visualized
with \texttt{plot(spar\_darwin,\ plot\_type\ =\ "coefs")}. In this data
set, the predictors are ordered by task, where the first 18 covariates
represent different features measured for the first task. Given that
there is clear grouping in the variables in this example, we can reorder
the coefficients for plotting by grouping them by feature, rather than
task. This allows to assess how the different features (e.g., time it
takes to complete a certain task) relate to the likelihood of having AD
and how stable the sign and magnitude of the coefficient is across the
models in the ensemble. We can achieve this by using reordering argument
\texttt{coef\_order} in method \texttt{plot} with
\texttt{plot\_type\ =\ "coefs"} (see Figure \ref{fig:darwin_coefs} which
can be reproduced with code in the supplementary materials).

\begin{figure}[t!]
\centering
\includegraphics[width=0.8\textwidth]{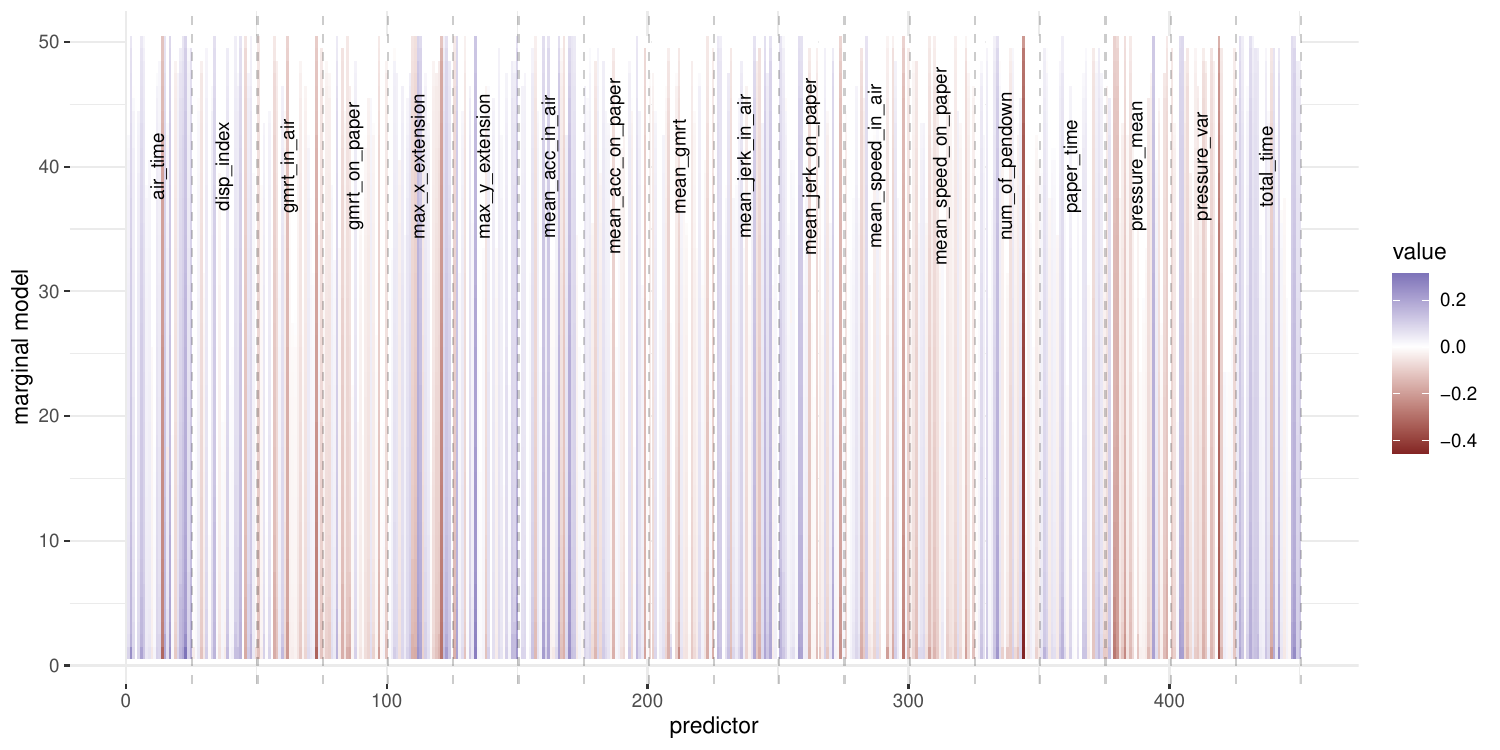}
\caption{Coefficient plot for all variables and all considered $M=50$ models in the SPAR ensemble
\emph{Darwin} data set. The coefficients are standardized, before thresholding. \label{fig:darwin_coefs}}
\end{figure}

In general we observe that the different features measures across
different tasks have the same impact on the probability of AD
(observable by the blocks of blue or red lines).

\section{Conclusion}\label{sec-conclusion}

Package \pkg{spar} can be employed for modeling data in a
high-dimensional setting, where the number of predictors is much higher
than the number of observations. The package provides an implementation
an algorithm for sparse projected and average regression (SPAR) proposed
in \citet{parzer2024glms} which combines variable screening and random
projection in an ensemble of GLMs. The package provides flexible classes
for i) specifying the coefficient based on which screening should be
performed (both in a classical fashion, where the predictors with the
highest screening coefficient are selected for subsequent or in a
probabilistic fashion, where variables are sampled for inclusion with
probabilities proportional to their screening coefficient), ii)
generating the random projection to be employed in each marginal model.
Screening coefficients based on marginal correlation between the
predictors and the response, marginal coefficients from a GLM or ridge
coefficients are provided in the package. Moreover, several random
projections are implemented: the Gaussian and sparse matrices which are
data-agnostic and satisfy the JL lemma and the data-driven projection
proposed in \citet{parzer2024sparse} for linear regression and extended
to GLMs in \citet{parzer2024glms}. This method has the advantage of
approximately capturing the true \(\beta\) in the span of the random
projection matrix, i.e., it ensures that the true regression
coefficients can be recovered approximately after the projection.
Methodologically, the SPAR algorithm, particularly when paired with the
data-driven random projection in \citet{parzer2024glms}, has been
demonstrated to perform effectively across different degrees of sparsity
of the coefficient vector and to offer competitive predictions and
variable ranking in both sparse and dense settings.

The flexibility and adaptability of the \pkg{spar} package make it an
attractive choice for practitioners and researchers. It encourages
exploration of new methods for variable screening and random projections
or the combination of existing approaches to tailor solutions to
specific data requirements.

\section*{Computational details}\label{computational-details}

The results in this paper were obtained using \proglang{R} 4.4.0.

\proglang{R} itself and all packages used are available from CRAN at
\url{https://CRAN.R-project.org/}.

\section*{Acknowledgments}\label{acknowledgments}

Roman Parzer and Laura Vana-Gür acknowledge funding from the Austrian
Science Fund (FWF) for the project ``High-dimensional statistical
learning: New methods to advance economic and sustainability policies''
(ZK 35), jointly carried out by WU Vienna University of Economics and
Business, Paris Lodron University Salzburg, TU Wien, and the Austrian
Institute of Economic Research (WIFO).

\renewcommand\refname{References}
  \bibliography{SPAR.bib}

\vfill

\end{document}